\documentclass[aps,prl,twocolumn,showpacs,amsmath,amssymb]{revtex4}
\usepackage{epsfig}
\usepackage{bm}

\newcommand{\AB}{{\it ab initio }}

\begin{document}

\title{TDDFT and Strongly Correlated Systems:
Insight From Numerical Studies}
\author{Claudio Verdozzi}
\affiliation{Mathematical Physics, Lund Institute of Technology, SE-22100 Lund, Sweden}
\date{\today}

\begin{abstract}
We illustrate the scope of Time Dependent Density Functional Theory (TDDFT) for strongly correlated (lattice) models out of equilibrium. Using the exact many body time evolution, we reverse engineer the exact exchange correlation (xc) potential $v_{xc}$ for small Hubbard chains exposed to time-dependent fields. We introduce an adiabatic local density approximation (ALDA) to $v_{xc}$ for the 1D Hubbard model and compare it to exact results, to gain insight about approximate xc potentials. Finally, we provide some remarks on the v-representability for the 1D Hubbard model.
\end{abstract}

\maketitle

\noindent Density Functional Theory (DFT) \cite{HoenKohn}
enables accurate investigations of realistic systems of considerable complexity.
However, strongly correlated systems (SCS) have until now remained elusive to DFT.
And, so far, most theories of SCS focus on equilibrium or non-equilibrium steady state regimes, to understand the long time response to external fields. 
Nanoscale systems pose new challenges to the theory of strong correlations, since
the latter are usually enhanced by spatial confinement. Virtually every future (nano) technology will use devices which interact 
with a time dependent (TD) environment. This increases the demand for
\AB methods to describe realistic SCS acted upon by fast TD external fields.

In the last decade, TDDFT \cite{Hardy1} has emerged as an effective
\AB treatment of TD phenomena \cite{TDDFTbook,botti,burke}. 
DFT and TDDFT functionals,  although related, are different entities \cite{Maitra}: 
progress within TDDFT comes with progress with non equilibrium functionals. 
Constructing TDDFT functionals is an active area of research, with much work done, for example, in terms of the so-called Optimised Effective Method and extensions \cite{Ullrich,Gorling}. A systematic route is given 
by a variational approach to Many Body Perturbation Theory (MBPT)  \cite{abl}, with
a controlled improvement of the functionals \cite{UvBNDRvLGS}. We also mention recent work  \cite{CapelleHooke} to include corrections to the ALDA \cite{Soven}. 
Current TDDFT functionals are quite successful for weakly interacting systems or in the linear response regime\cite{TDDFTbook,botti,burke}. 
To date, no studies are available of TDDFT applied to SCS (for DFT, see \cite{GunSchon,Capellea}). At this early stage, model systems can be of aid, to provide guidelines for \AB approaches. An assessment of TDDFT for model SCS under TD fields is thus highly desirable.

Here we study finite Hubbard chains in the presence of TD external fields, and use the results from exact time evolution to assess the potential of TDDFT for SCS. We also introduce an ALDA to $v_{xc}$, based on  an LDA-Bethe-Ansatz approach to the ground state of the inhomogeneous 1D Hubbard model \cite{Capellea}.
Our main results are i) in the range of parameters we investigated, 
TDDFT is a practically viable route to describe
SCS far away from equilibrium and in the TD regime; this is our central result; 
ii) an exact analytic treatment for a two-site chain and an exact inequality for
general 1D chains are consistent with the numerical results; 
iii) for not too large external fields, the exact xc potential, $v_{xc}$, 
obtained numerically by reverse engineering, is regular and well behaved within the time span
of our simulations. However, in some cases,  
$v_{xc}$ shows sharp structures in its temporal profile;
iv) strong electron-electron interactions reduce memory effects; yet, non-adiabatic and non-local effects are in general necessary  ingredients 
for a TDDFT of SCS.

%
%
%
%
\noindent {\it TDDFT time evolution for the many body problem.} 
%
%
%
We study open-ended Hubbard chains, with Hamiltonian
\begin{equation}
H\!= V\!\!\sum_{\langle RR'\rangle \sigma }a^{+}_{R\sigma}a_{R'\sigma}\!\!+U\sum_{R}n_{R+}n_{R-}
+h(t)\sum_{\sigma} n_{1\sigma}
\label{ham}
\end{equation}
with $n_{R\sigma}=a_{R\sigma}^{\dagger}a_{R\sigma}$, 
$\sigma=\uparrow,\downarrow$ and $\langle R,R'\rangle$
denoting nearest neighbour sites. The hopping parameter is $V$ ($V=-1$),
$U$ is the interaction strength, and $h(t)$ is the strength of a spin
independent, local external field. $U$ and $h(t)$ are given in units of $|V|$.
For simplicity, $h(t)$ is 
localised at the leftmost site ($R=1$), but we examined 
other couplings, not discussed here. We consider $L=4$ to $12$ sites and $N_{e}=L$ or $3L/2$ electrons (half- and three-quarter filling densities); we take spin up and down electrons equal in number; this holds during the time evolution, since $H$ has no spin-flip terms.
To evolve in time the exact many-body $|\Psi(t)\rangle$ we use 
the Lanczos's algorithm \cite{JChemphys} 
in the mid-point approximation (in all calculations the timestep $\Delta=0.02|V|^{-1}$;
numerical convergence was checked by halving $\Delta$). By a fitting procedure, we find an exact Kohn-Sham (KS) Hamiltonian,\newline
{\footnotesize $H^{KS}=V\sum_{\langle RR'\rangle \sigma }a^{+}_{R\sigma}a_{R'\sigma}+h(t)\sum_{\sigma}n_{1\sigma}+\sum_{R\sigma}v_{eff}(R,t) n_{R\sigma}$},\newline
where $v_{eff}(R,t)=v_H(R,t)+v_{xc}(R,t)$. 
Since we consider nonmagnetic regimes, $v_{eff}(R,t)$ is spin-independent.
We require the exact and the KS electron densities to be the
same at each time and cluster site. In practice, $v_{xc}(R,t)$ is determined by 
minimising $\sum_{R\sigma} [\langle n_{R\sigma}\rangle^{KS}_t - \langle n_{R\sigma}\rangle_t]^2$. Our procedure extends to TDDFT a method introduced long ago in DFT \cite{AlmbladhBarthreveng}. In \cite{vLPRL}, it was shown quite generally how to map  from TD densities to potentials. Such mapping was recently used for a He model atom \cite{exactwoparticle}.

%
\begin{figure}[tbp]
\includegraphics*[width=.47\textwidth]{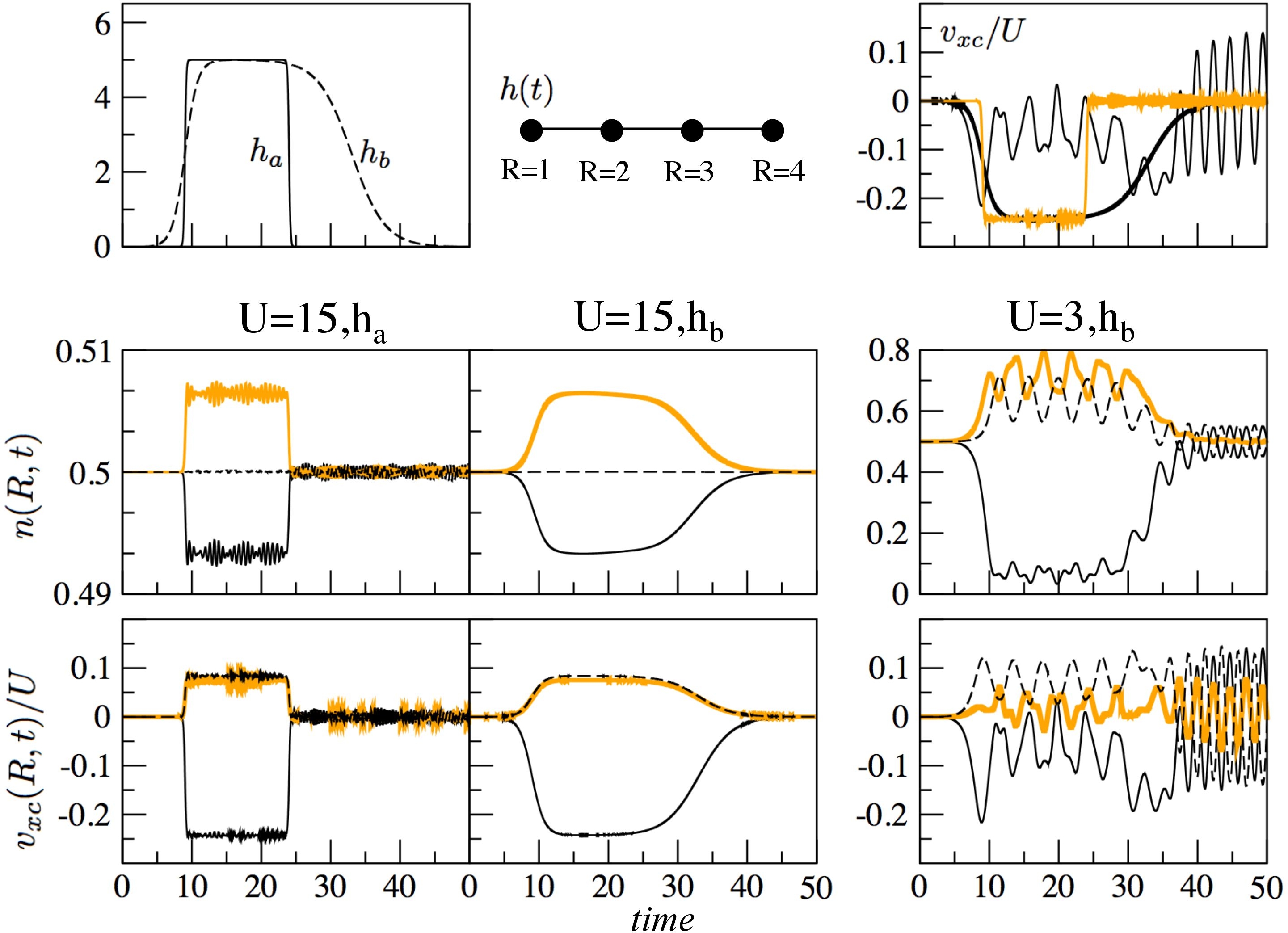}
\caption{ (Color online) $L=4, N_e=4$. Top left : the external fields 
$h(t)=h_{a,b}(t)$.  Mid and bottom rows: $n(R,t)$ and $v_{xc}(R,t) $ for
different ($U,h$) cases: black solid, orange solid (grey in b\&w) and black dashed curves
refer to sites $R=1,2,4$, respectively (R=3 results can be obtained via $\sum_R v_{xc}(R,t)=0,
 \sum_R n(R,t)=2$).
Top right: $v_{xc}/U$ at $R=1$ for
$U=3,h_b$ (thin black curve), $U=15,h_a$, (orange/grey curve),
$U=15, h_b$ (thick black curve). 
All panels have the same time intervals/units.}
\label{fig1}
\vspace{-0.5cm}
\end{figure}
%
\noindent {\it Half filled chains: Exact Results and TDDFT}.
Small clusters prevent local excitations from propagating away,
introducing time oscillations in the density; as a way
to mitigate size effects, we consider chains of different lengths.
We start with a four-site chain at half-filling (Fig. \ref{fig1} ). 
In the initial, ground state, $n(R,t=0)\equiv \langle g|n_{R\sigma}|g\rangle=0.5$ at any site. 
We choose $U=3$ and $15$, as examples of two interaction regimes.
The results for $v_{xc}$ are shown in the bottom panels: when $v_{xc}$ is
reused in the KS equations, it reproduces $n(R,t)\equiv\langle n_{R\sigma}\rangle_t$
(middle panels) with an accuracy of $10^{-5}$ or better (this applies to all
figures). Since $v_{xc}$ is defined up to an arbitrary site independent
function $C(t)$, we display
the potentials differences, e.g. $\delta v_{xc} (R,t)=v_{xc} (R,t)-v^{av}_{xc} (t)$,
where $v^{av}_{xc} (t)=(1/L)\sum_i v_{xc} (R_i,t)$. This also applies to 
$v_{H}$ and $v_{eff}$( for the Hartree term,  $v^{av}_H(t)=UN_e/2L$). 
For simplicity, the  prefix $\delta$ will be omitted.
Also, we find useful to rescale $v_{xc}$ ( and $v_{eff}$) by $U$ 
when comparing results for different $U$'s. 
In Fig. \ref{fig1} we consider two external perturbations, 
switched-on/off at a faster ($h_a$) or slower ($h_b$) rate. For  
$U=15$ and $h=h_a(t)$, the 
densities exhibit fast oscillations superimposed on a smoother, average change.
For $U=15$ and $h_b(t)$, the oscillations are considerably suppressed,
due to a more gradual change of the overlap between the initial, ground state
and the excited ones during the onset of $h_b(t)$.
The degree of charge redistribution is determined by $U$:
for example, for $h_b(t)$, the (small) charge imbalance at $U=15$
is fully absorbed by the second, $R=2$, site; at $U=3$,
all sites are involved. From a TDDFT perspective, this is
a consequence of how $v_{xc}$ depends on $U$. In the bottom panels, 
we can see that $v_{xc}$ and $n(R,t)$ behave rather similarly. 
On the other hand, results in the top right panel of Fig. \ref{fig1}
show that the range variation of the rescaled quantities, i.e. $v_{xc}/U$,
is comparable in all cases. Also, for large $U$,
$v_{xc}/U$ is very much in phase with the perturbation.
For $U=3$, out-of-phase effects are evident. For example, the
largest oscillations in $v_{xc}/U$ occur  when $h_b(t)$ has already returned to $0$
(this suggest that large $U$ values tend to reduce memory effects).
This is a rather generic behavior, that we noted also for other fields \cite{Verdozzi07}.\newline
%
\begin{figure}[tbp]
\includegraphics*[width=.48\textwidth]{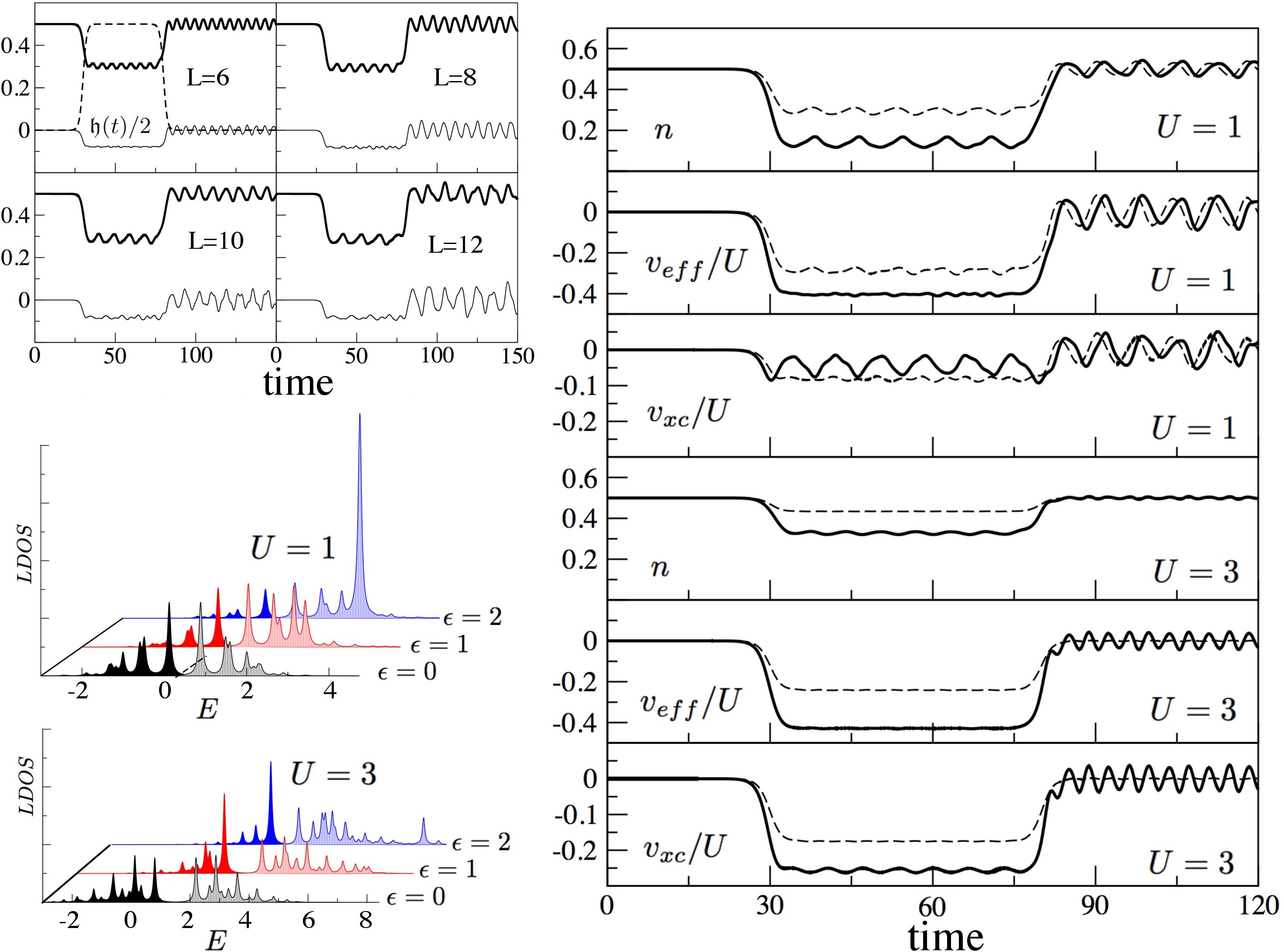}
\caption{
(Color online) Half-filling results. Top left four panels: exact density (thick black) and 
$v_{xc}/U$ (thin black) curves at site $R=1$ when $U=1$ and L=6 to 12. The field $\mathfrak h(t)$ is also shown (dashed curve).
Right six panels:  $L=8$. 
Dashed (solid) curves
refer to $h_1(t)=\mathfrak h(t)$ ($h_2(t)=2\mathfrak h(t)$). 
Left bottom graphs: ground state LDOS  at $R=1$ for $L=8$,
with a site energy shift $\epsilon_{R=1}=0, 1, 2$. Darker (lighter) patterns refer to the hole (electron)
LDOS. A Lorentzian broadening was introduced.}
\label{fig3}
\vspace{-0.5cm}
\end{figure}
%
\noindent {\it Larger chains at half filling.}  In Fig. \ref{fig3}, top left panel, we show results for L=6- to 12-site clusters,
with $U=1$ and $h(t)=\mathfrak h(t)$ the same for all $L$'s.
Chains with different $L$ behave rather similarly, the obvious differences
being due to the fine details of the excited states.
In Fig. \ref{fig3}, in the six panels on the right,  we compare results for
$U=1$ and $3$ and two perturbations $h_1(t)=\mathfrak h(t)$ and $h_2(t)=2\mathfrak h(t)$,
when $L=8$. To get an idea of the strength of $h_{1,2}(t)$, we can look 
at the equilibrium one-particle interacting local density of states, LDOS (Fig. \ref{fig3}, bottom left panels), when a static shift $\epsilon_{R=1}=1,2$ is introduced.
The shift corresponds to the maximum value $h^{max}_{1,2}$ achieved by $h_{1,2}(t)$ 
during the time evolution (the unshifted LDOS is also shown) and
induces significant (larger for smaller $U$) 
spectral changes. The maxima
$h^{max}_{1,2}=1,2$ are large enough to induce transitions from occupied to empty levels
(cfr. with the energy gap in the unshifted LDOS). Results for the TD density at $U=1$ and $U=3$ in the six right panels are consistent with the LDOS features and with results from Fig. \ref{fig1}.
For example, charge variations are affected both by 
$U$ and $h(t)$: a larger $U$ (a smaller field) induces a weaker response.
Also, at $U=3$, the TD density imbalance is localised
near the site $R=1$; for U=1, it redistributes across all sites. 
Finally, in a broad parameter range, TDDFT
reproduces the exact density.\newline\
\noindent {\it Adiabaticity vs locality and TDDFT.} We now introduce an ALDA to TDDFT for the Hubbard model, and apply it to a chain with $L= 8$  and $N_e=12$ (thus we also show how TDDFT 
performs away from half filling).
To disentangle adiabatic from locality effects in $v_{xc}$, we use two
approximations (A1 and A2) for the density. In A1, we calculate at every timestep the ground state
one-particle density of  the instantaneous many body Hamiltonian, Eq.(\ref{ham}) (this implies no approximations based on local potentials).
In A2, we introduce an ALDA to $v_{xc}$; our ALDA uses a 
a Local Density Approximation (LDA) to $v_{xc}$ for  the ground state of the 1D inhomogenous Hubbard 
model \cite{Capellea}, based on the Bethe-Ansatz  (BA). 
We employ an analytical interpolation to $v^{{\scriptsize BA-LDA}}_{xc}(n)$ \cite{Capellea}:
\begin{eqnarray}
v^{{\scriptsize BALDA}}_{xc}(n) =\mu \left[2\cos(\pi z /\beta)-2\cos(\pi z /2)+Uz/2\right]
\label{BALDA}
\end{eqnarray}
where $n=n_{+}+n_{-}$, $z=1-|n-1|$, $\mu=sgn(n-1)$ and $\beta\equiv \beta(U)$, which is independent of $n$, 
is obtained from the BA solution at half filling \cite{Capellea}. 
From Eq.(\ref{BALDA}), $v_{xc}$ has  a jump at $n=1$, $\Delta_{xc}=4\cos(\pi/\beta(U))+U$.
In our novel ALDA scheme, $v^{{\scriptsize BALDA}}_{xc}$ becomes a function
of the instantaneous densities along the KS trajectories, $v^{{\scriptsize A2}}_{xc}\equiv v^{{\scriptsize BALDA}}_{xc}(n^{KS}(R,t))$.
In Fig. \ref{fig4}, we compare exact and approximate results for slow and 
fast perturbations.
All results are for site $R=1$. We begin with the non-adiabatic case (panels in the top four rows) where $U=3,6, h_{1}(t)=\mathfrak h(t)$ and $ h_{2}(t)=2 \mathfrak h(t)$, with $\mathfrak h(t)$ the same as in 
Fig. \ref{fig3}.
An exact TDDFT description (black solid curves)
is possible also away from half filling $n>1$. As when $n=1$, but to a lesser extent, larger $U$ 
values reduce the changes in the density due to $h(t)$.
For $h_2(t)$ and $U=3$, the exact $v_{xc}$ exhibits
sharp resonances for $t>90$, when $h_2(t)$ has returned to zero.
At the same points, the density behaves smoothly.
We observed such peaks for other kinds of perturbations 
and other parameters values; their intensity increases at larger
perturbation strengths. Such structures might
be a challenge in constructing approximate potentials. 
Turning to A1, we note 
that a non-local but fully adiabatic description (dashed curves) gives correctly
the average profile of the density. With the exact wavefunction replaced by
the instantaneous (exact) ground state counterpart, there is no contribution from the
excited states. This removes memory effects, and the density closely follows
the temporal profile of  $h(t)$: for example, the oscillations in the exact densities around $t=90$ are completely missed by A1. 
For A2, (orange curves, grey in b\&w) the agreement with the exact densities is better, the maximum discrepancy being within a few percent (this level
of discrepancy we also found for BALDA ground state densities). When $h_{1,2}(t)=0)$, A2 reproduces many aspects of the exact results. However, a significant time-delay of
certain traits suggests that memory effects are not being properly taken into account (this especially manifests for $t\ge90$, when $h_{1,2}(t)=0)$. 
The agreement between exact and A2 densities looks better for $h_2(t)$ than  for
$h_1(t)$. To elaborate on this point, we compare exact and A2 results for $v_{xc}$.
For $h_1(t)$ and both $U$ values, there is a small discrepancy between $v_{xc}$ and $v^{\scriptsize{A2}}_{xc}$. This accounts for part of the difference between the corresponding densities (the other part being due to $v_{H}$). 
For $h_2(t)$,  when $2n^{\scriptsize{KS}}(R,t)=1$,   $v^{\scriptsize{A2}}_{xc}$ shows discontinuities
(see Eq.(\ref{BALDA})) which are absent in $v_{xc}(t)$. 
This suggests that,
for $h_2(t)$,  the agreement between A2 and exact results is somewhat
accidental, whenever $h(t)$ drives $n$
across the half-filling value. To corroborate this point, we show 
(Fig. \ref{fig4}, bottom row) results for two slow
perturbations $h^a_{1}(t)=\mathfrak h_{ad.}(t)$ (dashed curves) and 
$h^a_2(t)=2\mathfrak h_{ad.}(t)$ (solid curves), with $\mathfrak h_{ad.}(t)$
a smoothened version of $\mathfrak h(t)$ of Fig. \ref{fig3}. 
The exact (black curves) and A1 densities are identical (A1 densities are fully underneath the exact ones), i.e. the exact time evolution is fully adiabatic; A2 performs well (orange curves, grey in b\&w)
whenever $n^{\scriptsize{KS}}(t)$ does not cross 
the half filling point. If the crossing occurs ($U=3, h^a_2(t)$), we see noise-like features
in $n^{\scriptsize{KS}}$, due to the jump in $v^{A2}_{xc}$. 
To summarize this section, TDDFT reproduces the exact density 
with a reasonable-looking $v_{xc}$ which can be, however,
rather different from the ALDA one. And, in general,
non-adiabatic and non-local effects (the jump is a non local feature)
are both needed in $v_{xc}$ for a TDDFT of SCS. 
We finally note that for SCS with nearly filled bands 
the T-matrix approximation \cite{Tmatrix0}, TMA, 
is quite successful \cite{Tmatrix8}. As a mention of work in progress,
a study of $v_{xc}$ in the TD TMA is under way.
\newline
%
\begin{figure}[tbp]
\includegraphics*[width=.48\textwidth]{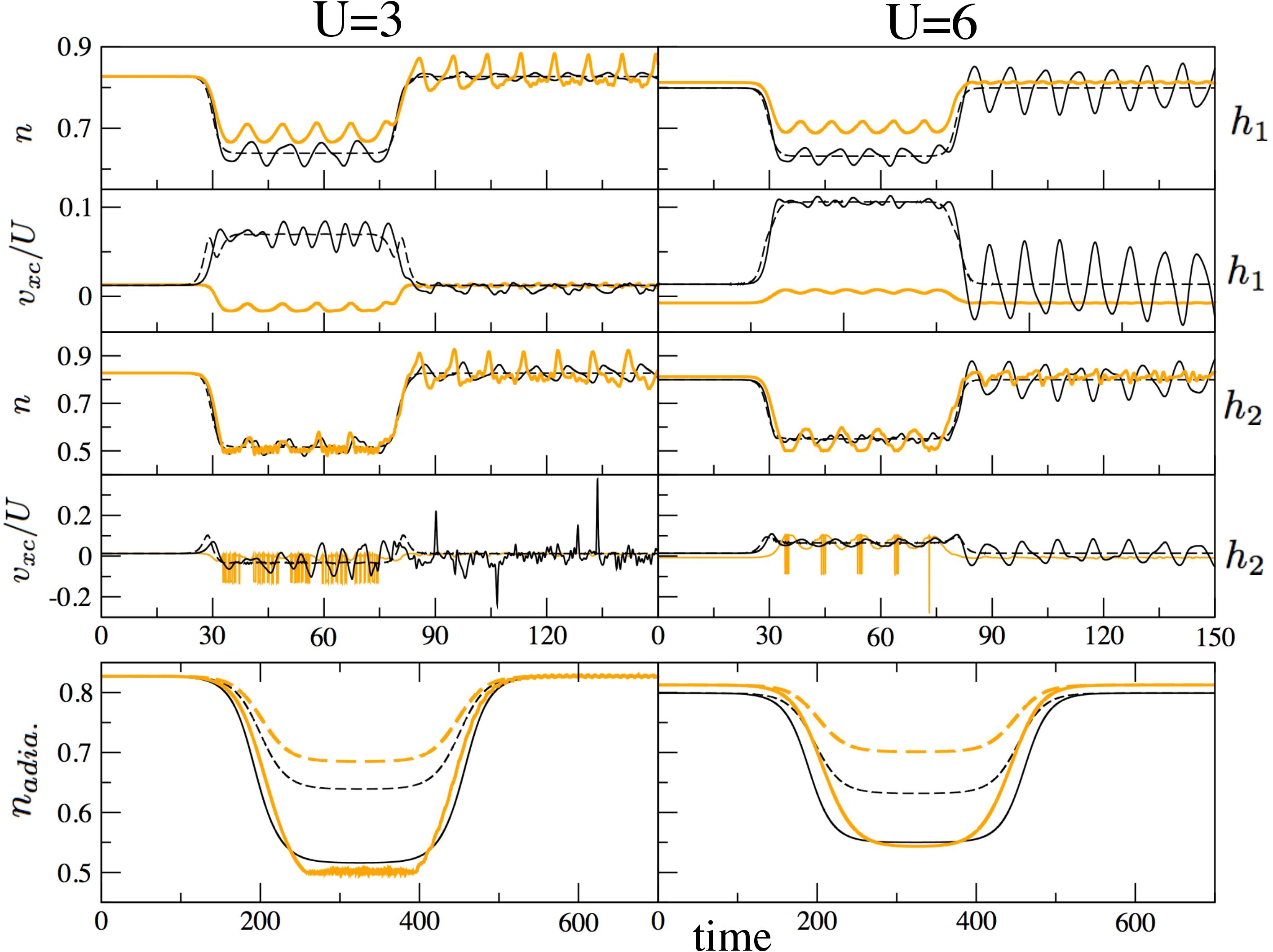}
\caption{(Color online) L=8 at 3/4 filling. Top four rows: fast fields.
Black solid, black dashed, and orange (grey in b\&w) curves denote 
exact, A1 and A2 results, respectively. The time unit is the same for all panels; 
panels in the same row share the same vertical scale.
The curves in each row (i.e. $n_1,v_{xc}/U$) are specified on the left, while the
fields (i.e. $h_1$ or $h_2$ ) are specified on the right. Bottom row:  slow fields (note
the horizontal axis). Solid (dashed) curves
refer to $h^a_1$ ($h^a_2$) fields. Black (orange/grey) curves denote exact (A2) results.}
\label{fig4}
\vspace{-0.5cm}
\end{figure}
%
\noindent {\it Remarks about v-representability.} It was recently pointed out \cite{Baer} that, for lattice models, there is an issue concerning the mapping of TD densities to potentials. 
The analysis in \cite{Baer} was for a one-particle, two site system.
To discuss the Kohn-Sham v-representability (KSVR) in a two-site system with $N_e=1$ electrons, 
we write, for any time $t$,
$\psi_{KS}(t)=e^{i\chi}{\scriptsize \left( n^{1/2},  e^{i\phi}(1-n)^{1/2}\right)}$; one can show that $| \dot{n}| \le  2 |V| (n(1-n))^{1/2}$, a necessary and sufficient condition for KSVR. For the trial density in \cite{Baer}, there are time intervals where such inequality is not fulfilled. Similar conclusions for $L=2$ were independently reached
in \cite{Carsten}, where the $L>2, N_e=1$ was also
studied. In \cite{Carsten},
the inequality for $|\dot{n}|$ was related to the real vs complex nature of the effective potential. For $L>2, N_e=1$ the condition for a real potential is \cite{Carsten} $ |S_k|\le 2|V|\sqrt{n_k n_{k+1}}|$, where $S_k=\sum_{i=1}^{k}\dot{n_i}$.
However, for the present work, we need to consider the many-particle, interacting case. Starting with a Hubbard dimer (HD), i.e. with Eq.(\ref{ham}) for $L=2$, we  performed simulations for different pairs $(U, h(t))$ and verified that $|\dot{n}| \le  2 |V| (n(1-n))^{1/2}$ is always obeyed. This offers strong evidence of the KSVR of a HD. To complete a formal proof, one needs to show that such inequality always holds for an HD. This is indeed the case \cite{COA}.
The many-particle  $L>2$  case is considerably more complicated, and here we limit our discussion
to a simple but necessary condition for KSVR. For the KS system, the 
total density (per spin channel) at the $k$-th site is $n^{tot}_{k}=\sum_{\lambda}n^{\lambda}_k$, 
where $\lambda$  labels the KS one particle states. As a generalisation of the result in \cite{Carsten},  we define $S^{tot}_k=\sum_{\lambda}S^{\lambda}_k  = \sum_{i=1}^k \dot{n}^{tot}_i$.
We get $|S^{tot}_k| \le \sum_{\lambda} 2|V| (n^{\lambda}_{k} n^{\lambda}_{k+1})^{1/2}$
and, using the Schwarz inequality, $S^{tot}_k \le 2|V| (n^{tot}_{k} n^{tot}_{k+1})^{1/2} $.
For the interacting many body system, we start with  $ \langle \dot{n}_{ks}\rangle =i \langle [H,n_{ks}] \rangle
=iV\langle  [a^\dagger_{k+1,s}a_{k,s}-a^\dagger_{k,s}a_{k-1,s}-h.c.] \rangle$. We then get 
$S^{tot}_{k}=\sum_{i=1}^k  \langle \dot{n}_{ks}\rangle= iV \langle [a^\dagger_{k+1,s}a_{k,s}-h.c.] \rangle$.
By the same manipulations as in the HD, $|S^{tot}_k|\le 2|V| (n^{tot}_{k} n^{tot}_{k+1})^{1/2}$.
Thus, for $L>2, N_e>1$, the inequality holds for the KS and the interacting 1D systems, which is
consistent with the numerical results for $L>2$. \newline
%
In conclusion, we provided a characterisation of TDDFT for strongly correlated systems.
We compared exact vs. approximate results from the time evolution of model finite systems, in 
a broad range of model parameters.
The exact $v_{xc}$ gave us insight into some of the properties approximate xc functionals should satisfy. The v-representability problem was discussed, and an adiabatic approximation was introduced \cite{Polini}. Our results illustrate the scope of TDDFT for non equilibrium phenomena in the presence of strong, time varying external fields in SCS, and encourage further investigations, some of which currently under way. We acknowledge many profitable
discussions with C-O. Almbladh and U. von Barth. We also thank 
K. Capelle and K. Burke for useful conversations. This work was supported by the
EU 6th framework Network of Excellence
NANOQUANTA (NMP4-CT-2004-500198).

\begin{thebibliography}{02}
%
\bibitem{HoenKohn}
P. Hohenberg and W. Kohn, Phys. Rev. {\bf 136}, B 864 (1964);
W. Kohn and L.J. Sham, Phys. Rev.  {\bf 140}, A 1133 (1965).
%
\bibitem{Hardy1} 
E. Runge and E. K. U. Gross,  Phys. Rev. Lett. {\bf 52}, 997 (1984).
%
\bibitem{TDDFTbook}
{\it Time-Dependent Density Functional Theory}, edited by 
M.A.L. Marques, C. A. Ullrich, F. Nogueira, A. Rubio, K. Burke, E.K.U. Gross  
(Springer Verlag, 2006)
%
\bibitem{botti} 
S. Botti {\em et al.}, Rep. Prog.  Phys {\bf 70} 357(2007)
%
\bibitem{burke} 
K. Burke, J. Werschnik, E.K. U. Gross
J. Chem. Phys. {\bf 123} , 062206 (2005) 
%
\bibitem{Maitra}  N. T. Maitra, K. Burke and C. Woodward, Phys. Rev. Lett. {\bf 89}, 023002 (2002)
%
\bibitem{Ullrich} C. A. Ullrich, U. J. Gossmann, E.K.U. Gross, Phys. Rev. Lett. {\bf 74}, 872 (1995)
%
\bibitem{Gorling}  A. Gorling, Phys. Rev. A {\bf 55}, 2630 (1997)
%
\bibitem{abl} 
C.-O. Almbladh, U. von Barth and R. van Leeuwen, Int. J. Mod. Phys. B {\bf 13}, 535 (1999) 
%
\bibitem{UvBNDRvLGS} 
U. von Barth {\em et al.}, Phys. Rev. B {\bf 72}, 235109 (2005)
%
\bibitem{CapelleHooke}  E. Orestes {\em et al.},
J. Chem. Phys. {\bf 127}, 124101 (2007)
%
\bibitem{Soven}
A. Zangwill and P. Soven, Phys. Rev. A {\bf 21}, 1561 (1980) 
%
\bibitem{GunSchon}
K. Sch\"{o}nhammer, O. Gunnarsson, R.M. Noack, Phys. Rev. B{\bf 52}, 2504 (1995)
%
\bibitem{Capellea}
N. A. Lima {\em et al.}, Phys. Rev. Lett. {\bf 90} 146402 (2003)
%
\bibitem{Verdozzi07} For additional results, see C. Verdozzi, arXiv:0707.2317
%
\bibitem{JChemphys} T. J. Park and J. C. Light, J. Chem. Phys. 85, {\bf 10}, 5870 (1986)
%
\bibitem{AlmbladhBarthreveng} 
C. O. Almbladh  and A. C. Pedroza, Phys.Rev.A  {\bf 29} 2322 (1984);  
U. von Barth,  in {\it Many Body Phenomena at Surfaces}, D. Langreth and H. Suhl eds., 
Academic Press (1984)
%
\bibitem{vLPRL} R. vanLeeuwen, Phys. Rev. Lett. {\bf 82}, 3863 (1999)
%
\bibitem{exactwoparticle} M. Lein and S. Kummel, Phys. Rev. Lett. {\bf 94}, 143003 (2005)
%
\bibitem{Tmatrix0} V. Galitzkii, Soviet Phys. JETP {\bf 7}, 104 (1958)%
%
\bibitem{Tmatrix8}
See, e.g., C. Verdozzi,  R. W. Godby,  S. Holloway,  Phy.  Rev.  Lett. {\bf 74}, 2327 (1995);
M. Cini and C. Verdozzi, Solid State Comm.  {\bf 57}, 657(1986)  
%
\bibitem{Baer} R. Baer,  J. Chem. Phys. {\bf 128}, 044103 (2008)
%
\bibitem{Carsten} Y. Li and C. A. Ullrich,  J. Chem. Phys. {\bf 129}, 044105 (2008)
%
\bibitem{COA} For a HD with two electrons with opposite spins, 
the inequality was proven by C.-O.Almbladh: 
one has ($V=-1$)
$\dot{\hat{n}}_s=i[ \hat{H}, \hat{n}_s] = i (a^{\dagger}_{2s}a_{1s}-
a^{\dagger}_{1s}a_{2s})$, where $s$ labels the spin channel ( $\langle \hat{n}_s \rangle= \langle \hat{n}_{-s} \rangle=n)$.  
Hence,  $|\langle \dot{\hat{n}}_s \rangle |= |  \langle a^{\dagger}_{2s}a_{1s} \rangle- \langle a^{\dagger}_{1s}a_{2s}\rangle| \le 2 | \langle a^{\dagger}_{1s}a_{2s}\rangle|$.
Using the Schwarz inequality, $2 | \langle a^{\dagger}_{1s}a_{2s}\rangle| \le 
2 | \langle a^{\dagger}_{1s}a_{1s}\rangle|^{1/2} | \langle a^{\dagger}_{2s}a_{2s}\rangle| ^{1/2}=
2 (n(1-n))^{1/2}$, where $n_1=n$ and $n_2=1-n$. 
%
\bibitem{Polini} Very recently, the ALDA introduced here (see also arXiv:0707.2317) has been used by 
W. Li, G. Xianlong, C. Kollath and M. Polini, arXiv:0805.4743
\end{thebibliography}
\end{document}